\documentclass[aps,prd,twocolumn,nofootinbib,showpacs]{revtex4}

\usepackage{graphicx,color}
\usepackage[colorlinks=true, linkcolor=blue, citecolor=blue, urlcolor=blue]{hyperref}

\begin{document}

\title{A solution to the $\gamma\gamma^*\rightarrow\eta_c$ puzzle using the Principle of Maximum Conformality}

\author{Sheng-Quan Wang$^{1,2}$}
\email[email:]{sqwang@cqu.edu.cn}

\author{Xing-Gang Wu$^2$}
\email[email:]{wuxg@cqu.edu.cn}

\author{Wen-Long Sang$^{3}$}
\email[email:]{wlsang@ihep.ac.cn}

\author{Stanley J. Brodsky$^4$}
\email[email:]{sjbth@slac.stanford.edu}

\address{$^1$Department of Physics, Guizhou Minzu University, Guiyang 550025, P.R. China}
\address{$^2$Department of Physics, Chongqing University, Chongqing 401331, P.R. China}
\address{$^3$School of Physical Science and Technology, Southwest University, Chongqing 400700, P.R. China}
\address{$^4$SLAC National Accelerator Laboratory, Stanford University, Stanford, California 94039, USA}

\date{\today}

\begin{abstract}

The next-to-next-to-leading order (NNLO) pQCD prediction for the $\gamma\gamma^* \to \eta_c$ form factor was evaluated in 2015 using nonrelativistic QCD (NRQCD). A strong discrepancy between the NRQCD prediction and the BaBar measurements was observed. Until now there has been no solution for this puzzle. In this paper, we present a NNLO analysis by applying the Principle of Maximum Conformality (PMC) to set the renormalization scale. By carefully dealing with the light-by-light diagrams at the NNLO level, the resulting high precision PMC prediction agrees with the BaBar measurements within errors, and the conventional renormalization scale uncertainty is eliminated. The PMC is consistent with all of the requirements of the renormalization group, including scheme-independence. The application of the PMC thus provides a rigorous solution for the $\gamma\gamma^* \to \eta_c$ form factor puzzle, emphasizing the importance of correct renormalization scale-setting. The results also support the applicability of NRQCD to hard exclusive processes involving charmonium.


\end{abstract}

\maketitle

Hadronic bound states of the heavy quarks, heavy quarkonia, with their rich display of dynamics at different mass scales, provide a unique platform for understanding Quantum Chromodynamics (QCD). Nonrelativistic QCD (NRQCD)~\cite{Bodwin:1994jh} has become the standard tool for studying the properties of heavy quarkonium as an effective field theory. Most measurements at the Tevatron, the B-factories, the LHC, etc., agree well with the NRQCD predictions, although some measurements have exposed serious inconsistencies, puzzles, and thus new challenges.

The simplest exclusive charmonium production process, $\gamma^*\gamma \to \eta_c$, measured in two-photon collisions, provides such kind of puzzle. One can define a transition form factor (TFF) $F(Q^{2})$ for charmonium production :
\begin{equation}
\langle \eta_c (p)\vert J_{\rm EM}^\mu \vert \gamma(k,\varepsilon) \rangle = i e^2 \epsilon^{\mu\nu\rho\sigma} \varepsilon_\nu q_{\rho} k_{\sigma} F(Q^2),
\end{equation}
where $J_{\rm EM}^\mu$ is the electromagnetic current evaluated at the time-like momentum transfer squared, $Q^2=-q^2 =-(p-k)^2>0$. In 2010, the BaBar collaboration measured the TFF $F(Q^{2})$ over the range of 2 GeV$^{2} < Q^{2} < 50$ GeV$^{2}$~\cite{Lees:2010de}; the measurements can be parameterized as $|F(Q^{2})/F(0)|=1/(1+Q^{2}/\Lambda)$ where $\Lambda=8.5\pm0.6\pm0.7$ GeV$^{2}$. The next-to-next-to-leading order (NNLO) NRQCD prediction fails to explain the BaBar measurements over a wide range of $Q^2$~\cite{Feng:2015uha}. As a consequence of this discrepancy, there has even been doubt whether NRQCD is applicable. However, a careful discussion of the renormalization scale dependence has been missing in the theoretical analysis. The renormalization scale was simply set as $\mu_r=\sqrt{Q^{2}+m_c^{2}}$, leading to a substantially negative NNLO contribution and a large value for $|F(Q^2)/F(0)|$, in disagreement with the data. This disagreement cannot be explained by taking higher Fock states into consideration~\cite{Guo:2011tz, Jia:2011ah}. A possible solution has been suggested by using a continuum approach to the two valence-body bound-state problem in relativistic quantum field theory~\cite{Raya:2016yuj, Chen:2016bpj}. In the present paper, we shall concentrate on how to solve the disagreement within the NRQCD framework.

The renormalization scale-setting problem is in fact an important issue for fixed-order pQCD predictions. Guessing the renormalization scale and setting an arbitrary range for its value introduces an arbitrary systematic error into pQCD predictions, which may lead to predictions inconsistent with the experimental measurements. The conventional treatment for setting the renormalization scale has a number of defects~\cite{Wu:2013ei}, such as:
\begin{itemize}
\item pQCD predictions based on a guessed scale are incorrect for the Abelian theory -- Quantum Electrodynamics (QED). In the case of QED, the renormalization scale of the running coupling $\alpha(q^2)$ reflects the virtuality $q^2$ of the photon propagator. It can be set unambiguously by using the Gell-Mann-Low procedure~\cite{GellMann:1954fq}, which automatically sums all proper and improper vacuum polarization contributions to the photon propagators to all orders;
 \item In contrast, predictions based on a guessed renormalization scale will depend on the renormalization scheme, violating a fundamental principle of the renormalization group;
 \item The perturbative series is in general factorially divergent at large orders like $n! \beta^n_0 \alpha_s^n$ -- the ``renormalon" problem~\cite{Beneke:1998ui};
 \end{itemize}
 It is often argued that the uncertainties in setting the renormalization scale will be suppressed by including enough high-order terms -- however, this expectation conflicts with the renormalon divergences. One then cannot decide whether the poor pQCD convergence is the intrinsic property of pQCD series, or is simply due to improper choice of scale. It is thus clearly important to reanalyze the NNLO prediction for the TFF $F(Q^{2})$ to clarify whether one is confronting the applicability of NRQCD or whether the discrepancy of experiment with theory is due to an improper choice of renormalization scale.

The Principle of Maximum Conformality (PMC)~\cite{Brodsky:2011ta, Brodsky:2012rj, Mojaza:2012mf, Brodsky:2013vpa} provides a systematic way to eliminate renormalization scheme-and-scale ambiguities of pQCD predictions. The PMC extends BLM scale-setting~\cite{Brodsky:1982gc} to all orders in $\alpha_s^n$. The PMC agrees with the first principles of QCD, which satisfies renormalization group invariance~\cite{Brodsky:2012ms, Wu:2014iba} and reduces in the $N_C\to 0$ Abelian limit~\cite{Brodsky:1997jk} to the standard Gell-Mann-Low method. In addition to pQCD predictions which are scheme independent, the number of active quark flavors $n_f$ appearing in virtual quark loops and the $\beta$-function is also correctly set by the PMC at each order. The PMC is thus an important advance for testing pQCD and other gauge theory predictions to high precision. There have been many important application of the PMC to collider processes. In each case the systematic error associated with renormalization scale uncertainties has been essentially eliminated~\cite{Wu:2015rga}. In this paper, we shall adopt the PMC to set the optimal renormalization scale for the TFF $F(Q^{2})$ by carefully dealing with the NNLO contributions.

The NRQCD prediction for $F(Q^{2})$ up to NNLO level, can be schematically written as
\begin{eqnarray}
F(Q^2)=c^{(0)}\left[1 + \delta^{(1)}a_s(\mu_r) + \delta^{(2)}(\mu_r)a^2_s(\mu_r)\right],
\end{eqnarray}
where $a_s(\mu_r)={\alpha_s(\mu_r)}/{\pi}$, and $\mu_r$ is the renormalization scale. The coefficient is
\begin{eqnarray}
c^{(0)}=\frac{4e^2_c\langle\eta_c|\psi^\dag\chi(\mu_\Lambda)|0\rangle}{(Q^2+4m^2_c)\sqrt{m_c}},
\end{eqnarray}
where $e_c=+2/3$ is the $c$-quark electric charge, $m_c$ is the $c$-quark mass, and $\langle\eta_c|\psi^\dag\chi(\mu_\Lambda)|0\rangle$ represents the nonpertubative input which can be eliminated by considering the ratio $|F(Q^2)/F(0)|$.

The NLO coefficient is $\delta^{(1)} = C_{F} f^{(1)}(\tau)$~\cite{Sang:2009jc} with
\begin{widetext}
\begin{eqnarray}
f^{(1)}(\tau) &=& \frac{\pi^2(3-\tau)}{6(4+\tau)}-\frac{20+9\tau}{4(2+\tau)}-\frac{\tau(8+3\tau)}{4(2+\tau)^2}\ln\left[\frac{4+\tau}{2}\right] +3\sqrt{\frac{\tau}{4+\tau}}{\rm ArcTanh}\left[\sqrt{\frac{\tau}{4+\tau}}\right] -\frac{\tau}{2(4+\tau)} \rm Li_2\left[-\frac{2+\tau}{2}\right] \nonumber\\
&&+ \frac{2-\tau}{4+\tau}{\rm ArcTanh}\left[\sqrt{\frac{\tau}{4+\tau}}\right]^2,
\end{eqnarray}
\end{widetext}
where $\tau\equiv Q^2/m^2_c$ and $C_{F}=4/3$. The NNLO coefficient $\delta^{(2)}$ can be divided into two parts, i.e., $\delta^{(2)}(\mu_r) = B_{\rm con}^{(2)}(\mu_r) + C_{n_f}^{(2)}(\mu_r)n_f$, where
\begin{eqnarray}
B_{\rm con}^{(2)}(\mu_r) &=& f^{(2)}_{\rm lbl}(\tau) + f^{(2)}_{\rm reg,con}(\tau) + \frac{11}{4}\ln\left[\frac{\mu^{2}_r}{Q^{2}+m^2_c}\right] \nonumber\\
&& C_F f^{(1)}(\tau) -\frac{\pi^2}{2}C_F(C_A+2C_F)\ln\left[\frac{\mu_\Lambda}{m_c}\right], \\
C_{n_f}^{(2)}(\mu_r) &=& f^{(2)}_{\rm reg,n_f}(\tau) - \frac{1}{6}C_F f^{(1)}(\tau) \ln\left[\frac{\mu^{2}_r}{Q^{2}+m^2_c}\right].
\end{eqnarray}
Here $C_A=3$ and $\mu_\Lambda$ is the factorization scale. The renormalization and factorization scale independent NNLO coefficients have been separated as the light-by-light contribution $f^{(2)}_{\rm lbl}(\tau)$ and the regular term $f^{(2)}_{\rm reg}(\tau)$ which is further divided as $ f^{(2)}_{\rm reg,con}(\tau)+f^{(2)}_{\rm reg,n_f}(\tau)n_f$. The light-by-light term $f^{(2)}_{\rm lbl}(\tau)$ is proportional to $n_f$; since it is free of ultraviolet divergences, it should be treated as a ``conformal" contribution when applying the PMC scale-setting. The NNLO coefficients at $Q^2=0$ are given explicitly by Feng et al.~\cite{Feng:2015uha}. We have recalculated the NNLO terms for arbitrary values of $Q^2$ and identified the $\beta$ contributions using the calculational methods described in detail in Ref.\cite{Feng:2015uha},

After applying the standard PMC procedures, the TFF $F(Q^{2})$ can be rewritten as
\begin{eqnarray}
F(Q^2)=c^{(0)}\left[1 + \delta^{(1)}a_s(\mu^{\rm PMC}_r) + \delta_{\rm con}^{(2)}(\mu_r)a^2_s(\mu^{\rm PMC}_r)\right],
\end{eqnarray}
where the PMC scale is
\begin{eqnarray}
\mu^{\rm PMC}_r=\mu_r\exp\left[\frac{3C^{(2)}_{n_f}(\mu_r)}{2 T_F \delta^{(1)}}\right],
\label{EQ:PMCscaleQ1}
\end{eqnarray}
and the conformal coefficient is
\begin{eqnarray}
\delta_{\rm con}^{(2)}(\mu_r)=\frac{11C_A}{2}C^{(2)}_{n_f}(\mu_r)+B^{(2)}_{\rm con}(\mu_r).
\end{eqnarray}
We have determined the PMC scale for the NLO-term by shifting the $\beta_0 = 11/3 C_A - 2/3 n_f$ term into the running coupling; the PMC scale for the NNLO-term is taken as the same scale as that of the NLO-term to ensure the renormalization scheme independence of $F(Q^2)$~\cite{Shen:2017pdu}.

For the numerical evaluation, we will take the input parameters as the same ones as those of Ref.\cite{Feng:2015uha}, e.g., the $c$-quark pole mass $m_c=1.68$ GeV and the factorization scale $\mu_\Lambda=m_c$. The running of the strong coupling is evaluated using the RunDec program~\cite{Chetyrkin:2000yt}.

\begin{figure}[htb]
\centering
\includegraphics[width=0.4\textwidth]{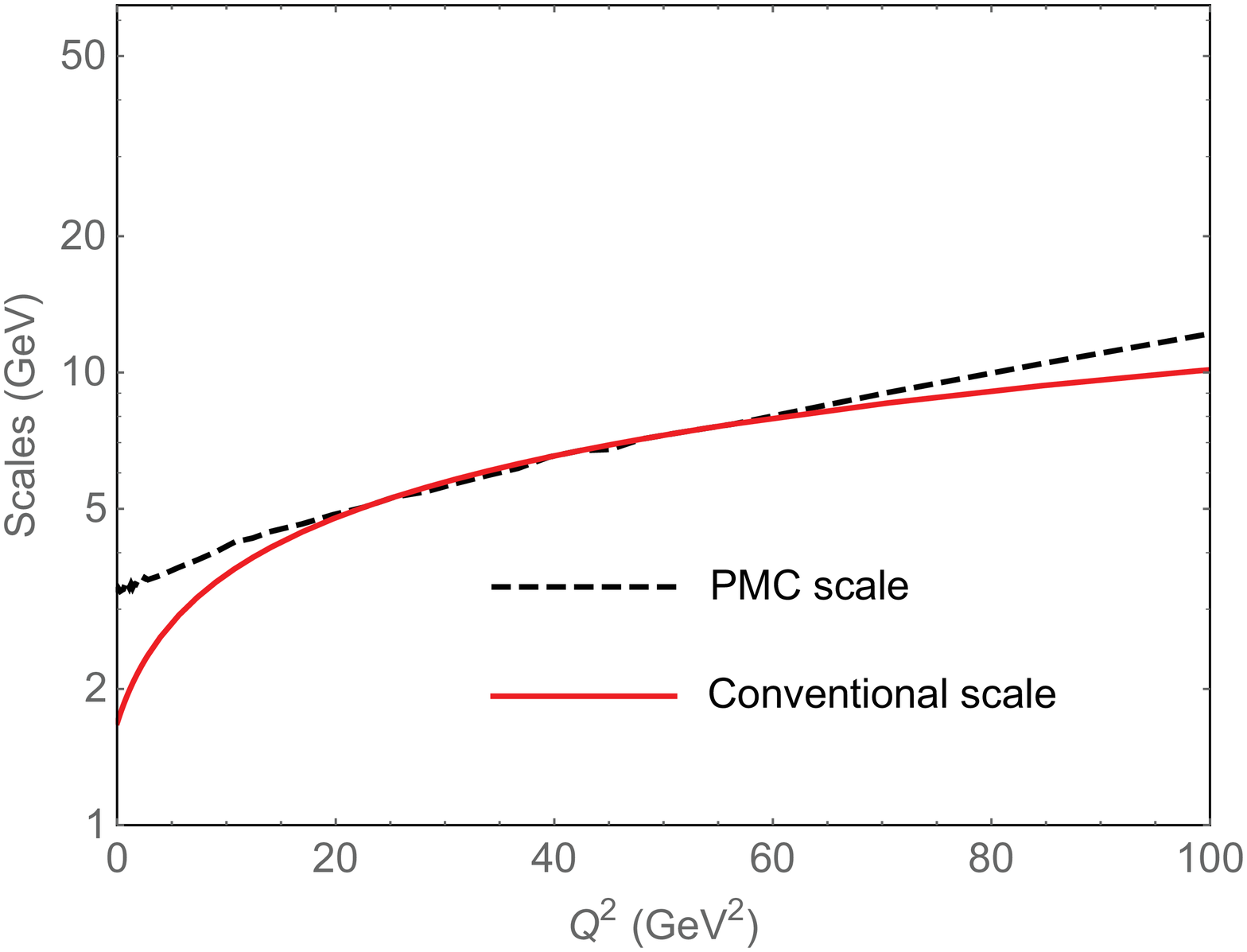}
\caption{The PMC scale of the NNLO TFF $F(Q^2)$ versus $Q^2$. The conventional choice of scale $\mu_r=\mu_Q$ is presented as a comparison. }
\label{FQscalePMC}
\end{figure}

The PMC scale $\mu^{\rm PMC}_r$ varies with momentum transfer squared $Q^2$ at which the TFF is measured. We present the PMC scale versus $Q^2$ in Fig.(\ref{FQscalePMC}), where the ``guessed" renormalization scale $\mu_Q=\sqrt{Q^2+m^2_c}$ is presented in comparison. The PMC scale $\mu^{\rm PMC}_r$ is larger than the ``guessed" value $\mu_Q$ at both the small and large $Q^2$-regions. The largest difference occurs at $Q^2=0$; in the intermediate $Q^2$-region: $Q^2\sim [20,60]$ GeV$^2$, the discrepancy between $\mu^{\rm PMC}_r$ and $\mu_Q$ is small.

When $Q^2=0$, the TFFs for three choices of $\mu_r$ under conventional scale-setting are
\begin{eqnarray}
F^{\rm Conv}(0)\left|_{\mu_r=1 {\rm GeV}}\right. &=& c^{(0)}(1 -0.25 -1.11). \\
F^{\rm Conv}(0)\left|_{\mu_r=m_c} \right.    &=& c^{(0)}(1 -0.18 -0.60). \\
F^{\rm Conv}(0)\left|_{\mu_r=2m_c} \right.    &=& c^{(0)}(1 -0.13 -0.36).
\end{eqnarray}
In the case of conventional scale-setting, the magnitude of the NLO contribution is moderate, whereas the NNLO-terms always give large negative contributions; the magnitude of the NNLO contribution is even larger than the LO terms for small $\mu_r$. In fact, if $\mu_r \lesssim 1.3$ GeV, the predicted value of $F^{\rm Conv}(0)$ becomes negative.

After applying the PMC, the PMC scale $\mu^{\rm PMC}_{r}$ in $\overline{\rm MS}$-scheme is fixed at $3.28$ GeV for any choice of the initial scale $\mu_r$ --  the strong scale dependence of conventional scale-setting is eliminated; i.e.,
\begin{equation}
F^{\rm PMC}(0) \equiv c^{(0)}(1-0.13-0.37)
\end{equation}
for any choice of $\mu_r$. It is interesting to find that if one assumes $\mu_r\simeq 2 m_c$, the pQCD prediction using conventional scale-setting is close to the PMC one. Thus the optimal choice for $F^{\rm Conv}(0)$ using conventional scale-setting is actually $2m_c$, rather than the conventional choice $m_c$.

\begin{figure}[htb]
\centering
\includegraphics[width=0.4\textwidth]{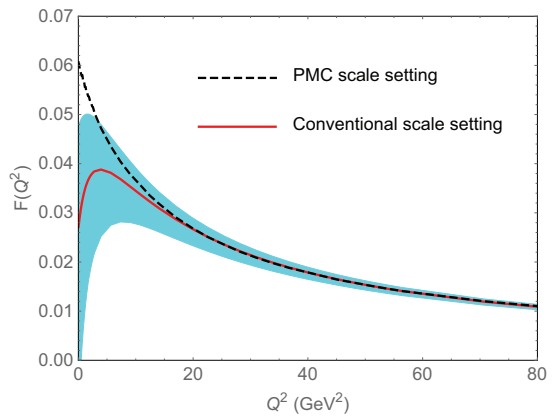}
\caption{The NNLO TFF $F(Q^2)$ versus $Q^2$ using conventional and PMC scale-settings. The shaded band shows the conventional scale uncertainty for $\mu^2_r=[\mu^2_Q/2, 2\mu^2_Q]$, whose central value (the solid line) is for $\mu_r=\mu_Q$. The PMC prediction (the dashed line) is independent of the choice of $\mu_r$. }
\label{FQF0FQOF0}
\end{figure}

The magnitude of the $n_f$-dependent term $C_{n_f}^{(2)}$ is small, and the NNLO conformal coefficient $\delta_{\rm con}^{(2)}(\mu_r)$ is close to the NNLO coefficient $\delta^{(2)}(\mu_r)$; e.g., at $Q^2=0$, $\delta_{\rm con}^{(2)}(\mu_r)=-61.2$, and $\delta^{(2)}(\mu_r)=-56.5^{+4.9}_{-3.7}$ for $\mu_r\in[1$ GeV, $2m_c]$. Thus the running behavior of the coupling constant determines the exact value of the pQCD series, identical to the value determined using the RGE and the PMC scale-setting. We present the NNLO TFF $F(Q^2)$ for conventional and PMC scale-setting in Fig.(\ref{FQF0FQOF0}). It shows that the PMC prediction is independent of the choice of $\mu_r$, whereas the conventional scale uncertainty is large, especially in low $Q^2$-region.

In addition to the renormalization scale uncertainty, there are other uncertainty sources, such as the nonperturbative matrix element, the factorization scale $\mu_\Lambda$, and the $c$-quark mass $m_c$. We will consider the ratio $|F(Q^2)/F(0)|$ in order to suppress the uncertainty from the nonperturbative matrix-element.

The factorization scale uncertainty exists even for conformal theories; this uncertainty can be resolved by matching the perturbative prediction with the non-perturbative bound-state dynamics~\cite{Brodsky:2014yha}.  In our case, we observe a very small factorization scale dependence after determining the optimal renormalization scale predicted by the PMC. For example, in the case $Q^2=0$, a large factorization scale uncertainty is observed using conventional scale-setting; i.e.,
\begin{equation}
F^{\rm Conv}(0)|_{\mu_r=m_c}=0.43c^{(0)},\;\; 0.22c^{(0)},\;\; -0.06c^{(0)}
\end{equation}
for $\mu_\Lambda=1$ GeV, $m_c$ and 2$m_c$, respectively. The magnitude of the negative NNLO term increases with increasing $\mu_\Lambda$, this explains why $F^{\rm Conv}(0)$ is negative for $\mu_\Lambda =2m_c$. On the other hand, by applying the PMC, we obtain a reasonable factorization scale dependence
\begin{equation}
F^{\rm PMC}(0)=0.61c^{(0)},\;\; 0.50c^{(0)},\;\; 0.34c^{(0)}.
\end{equation}

\begin{figure}[htb]
\centering
\includegraphics[width=0.4\textwidth]{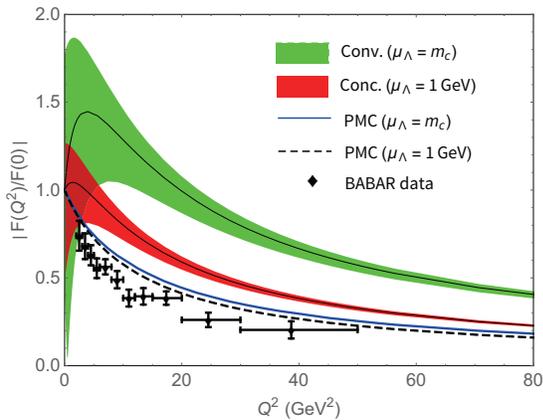}
\caption{The NNLO ratio $|F(Q^2)/F(0)|$ versus $Q^2$ using conventional (Conv.) and PMC scale-settings, where the BaBar data are presented as a comparison~\cite{Lees:2010de}. The predictions under conventional scale-setting are the same as those of Ref.\cite{Feng:2015uha}. The error bars are for $\mu^2_r=[\mu^2_Q/2, 2\mu^2_Q]$. $m_c=1.68$ GeV. }
\label{FQF0DataPMC168}
\end{figure}

We present a comparison of both the renormalization and factorization scale dependence on the NNLO ratio $|F(Q^2)/F(0)|$ using conventional and PMC scale-settings in Fig.(\ref{FQF0DataPMC168}). The predictions using conventional scale-setting agree with those of Ref.\cite{Feng:2015uha}. In the case of conventional scale-setting, choosing a smaller factorization scale could lower the NNLO ratio $|F(Q^2)/F(0)|$ to a certain degree, but it cannot eliminate the large discrepancy with the data. In contrast, the PMC prediction has a small factorization scale dependence and is close to the BaBar measurement~\cite{Lees:2010de}.

\begin{figure}[htb]
\centering
\includegraphics[width=0.4\textwidth]{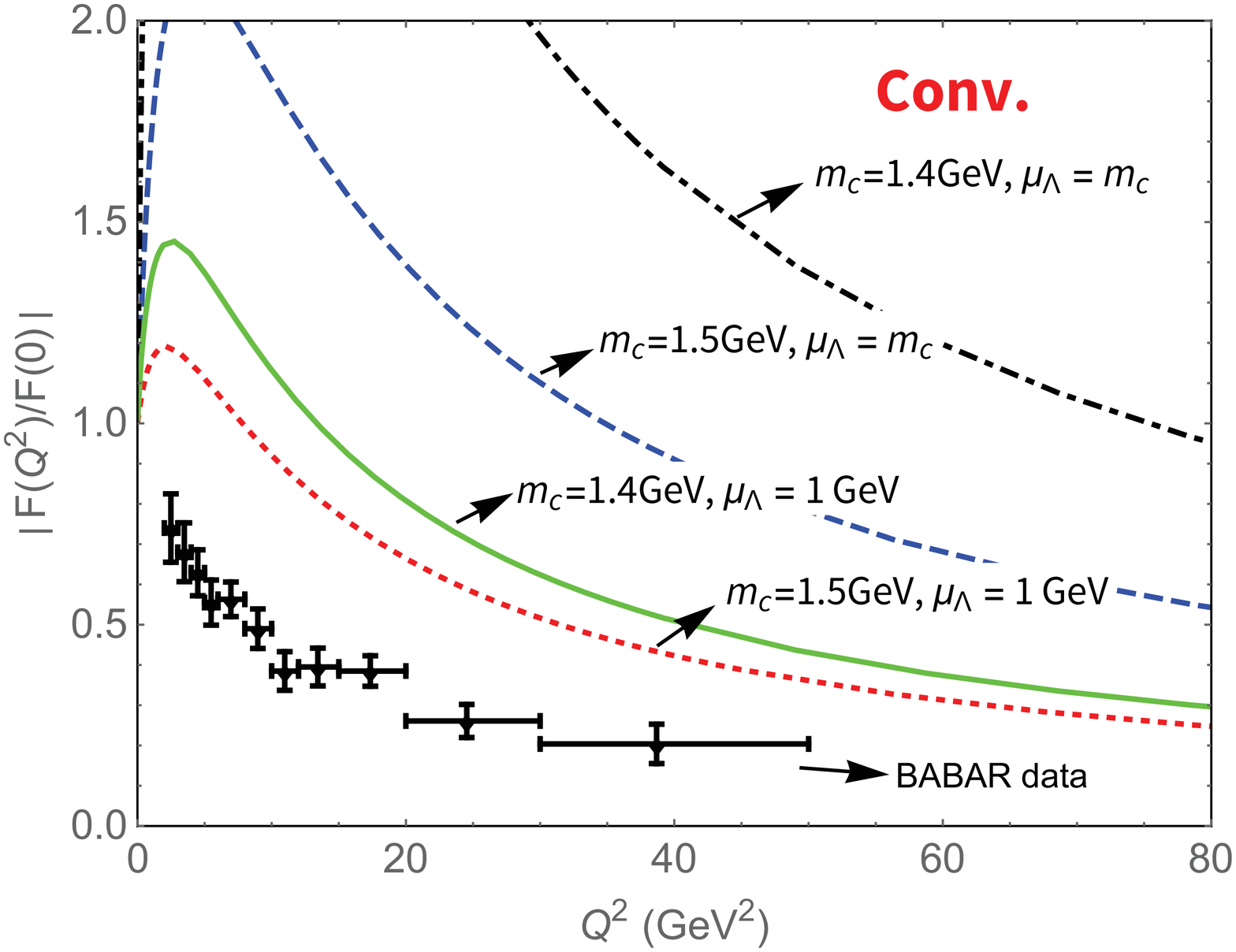}
\hspace{0.2cm}
\includegraphics[width=0.4\textwidth]{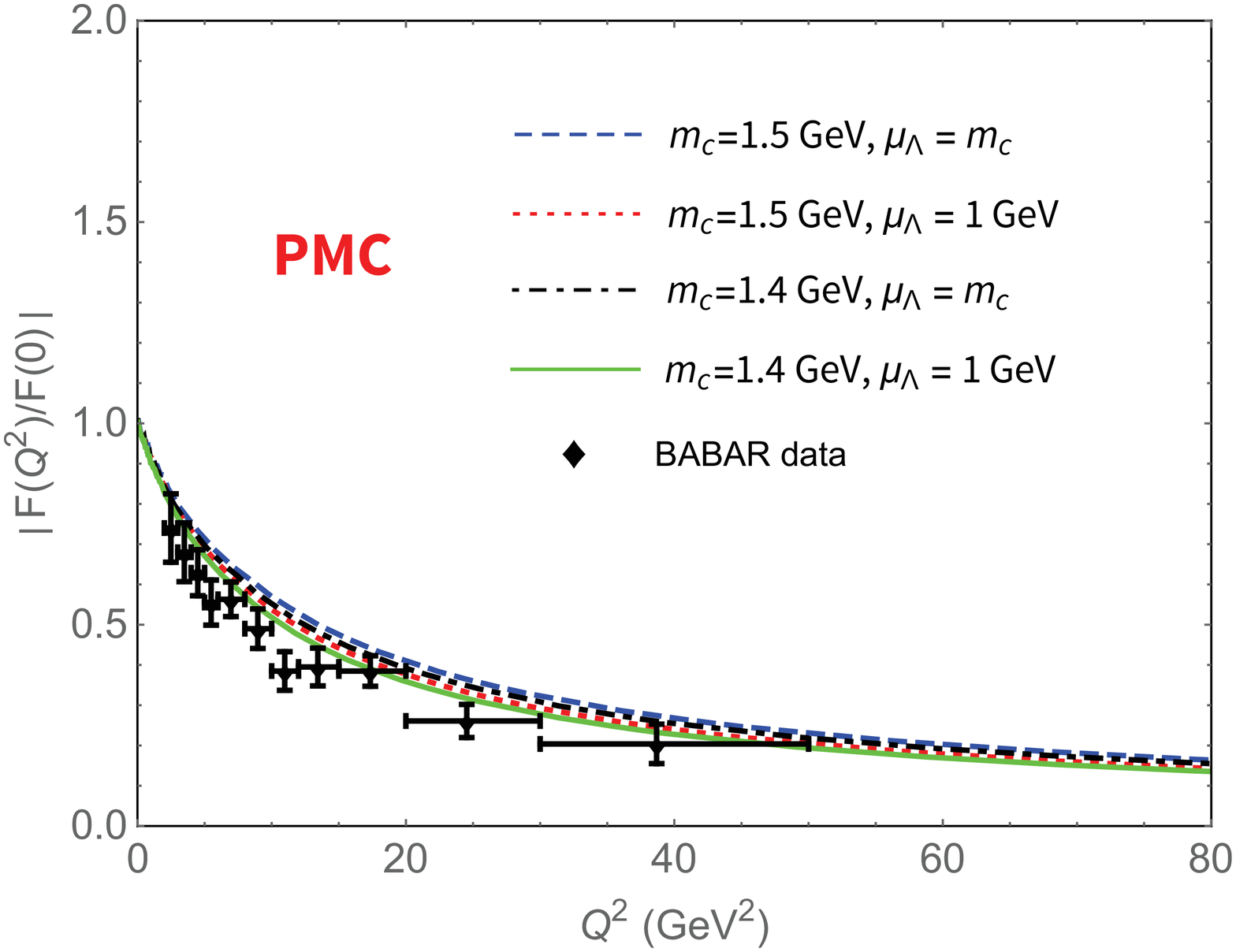}
\caption{The NNLO ratio $|F(Q^2)/F(0)|$ versus $Q^2$ using conventional (Up) and PMC (Down) scale-settings for different values for the quark mass $m_c$ and the factorization scale $\mu_\Lambda$. }
\label{FQF0DataPMC15}
\end{figure}

We also note that the dependence on the charm quark mass $m_c$ is large for conventional scale-setting. At $Q^2=0$, $F^{\rm Conv}(0)=0.22c^{(0)}$, $0.14c^{(0)}$, $0.07c^{(0)}$ for $m_c=1.68$, $1.5$, and $1.4$ GeV, respectively. However, the PMC prediction for the TFF displays a more reasonable dependence; i.e., $F^{\rm PMC}(0)=0.50c^{(0)}$, $0.46c^{(0)}$, $0.43c^{(0)}$ for $m_c=1.68$, $1.5$, $1.4$ GeV, respectively. We compare the NNLO ratio $|F(Q^2)/F(0)|$ using conventional and PMC scale-settings in Fig.(\ref{FQF0DataPMC15}). Fig.(\ref{FQF0DataPMC15}) shows that the $m_c$ uncertainty from the value of the charm quark mass is minimal for any value of $Q^2$.

\emph{Summary}: We have recalculated the TFF $F(Q^2)$ up to the NNLO level by applying PMC renormalization scale-setting. In contrast, the prediction for the TFF using conventional scale-setting encounters strong dependence on choice of the renormalization scale, the factorization scale, and the value of $m_c$. Even allowing for these large uncertainties, the NNLO NRQCD prediction for the ratio $|F(Q^2)/F(0)|$ still deviates strongly from the BaBar measurement.

We have shown that by using the PMC to set the optimal renormalization scale, the conventional renormalization scale uncertainty in the NRQCD prediction for the charmonium TFF is eliminated. The dependence of the PMC prediction on both the choice of the factorization scale and the value for $m_c$ is greatly suppressed. The result is a precise NRQCD prediction which is in excellent agreement with the BaBar measurement. This application of the PMC emphasizes the importance of correct renormalization scale-setting; it also supports the applicability of NRQCD to hard exclusive processes involving heavy quarkonium. \\

{\bf Acknowledgements}: This work was supported in part by the Natural Science Foundation of China under Grant No.11547010, No.11625520, No.11705033 and No.11605144; by the Project of Guizhou Provincial Department under Grant No.2016GZ42963, No.KY [2016]028 and No.KY[2017]067; and by the Department of Energy Contract No.DE-AC02-76SF00515. SLAC-PUB-17247.

\end{document}